\newcommand{\dnsat}{\delta n^\mathrm{sat}}
\newcommand{\p}{\partial}
\newcommand{\Pseed}{P_\mathrm{seed}}

\documentclass[aps,prl,reprint,groupedaddress,floatfix]{revtex4-1}

\usepackage{graphicx}

\begin{document}


\title{Interplay of Laser-Plasma Interactions and Inertial Fusion Hydrodynamics}


\author{D. J. Strozzi\email{strozzi2@llnl.gov}, D. S. Bailey,  P. Michel, L. Divol, S. M. Sepke, G. D. Kerbel, C. A. Thomas, J. E. Ralph, J. D. Moody, M. B. Schneider}
\affiliation{Lawrence Livermore National Laboratory, Livermore, CA 94551}


\date{\today}

\begin{abstract}
The effects of laser-plasma interactions (LPI) on the dynamics of inertial confinement fusion hohlraums is investigated via a new approach that self-consistently couples reduced LPI models into radiation-hydrodynamics numerical codes. The interplay between hydrodynamics and LPI -- specifically stimulated Raman scatter (SRS) and crossed-beam energy transfer (CBET) -- mostly occurs via momentum and energy deposition into Langmuir and ion acoustic waves.  This spatially redistributes energy coupling to the target, which affects the background plasma conditions and thus modifies laser propagation. This model shows reduced CBET, and significant laser energy depletion by Langmuir waves, which reduce the discrepancy between modeling and data from hohlraum experiments on wall x-ray emission and capsule implosion shape.
\end{abstract}

\pacs{52.35.Fp, 52.38.-r, 52.38.Bv, 52.57.-z}

\maketitle

Interaction of a large-amplitude wave with other waves and background plasma is ubiquitous in plasma physics.  Magnetic fusion devices rely on heating and current drive by externally-launched RF, e.g.\ microwaves \cite{fisch-currdrive-rmp-1987}.  Their parametric decay has been studied since the 1970s \cite{porkolab-parametric-pof-1977}. Space plasmas show important interplay of waves and energetic particles, and ionosphere modification experiments by radar transmitters have shown anomalous absorption by decay into Langmuir waves (LWs) \cite{dubois-ionosphere-pop-2001}.  This Letter focuses on laboratory LPI, specifically in inertial confinement fusion (ICF), but is also relevant to parametric-amplifier \cite{malkin-ramanamp-prl-1999} and beam-combining schemes.

ICF entails compressing thermonuclear fuel (namely deuterium and tritium) to 1000 g/cm$^3$ and $\sim$10 keV temperature. Most research uses lasers to implode a low-$Z$ capsule either by direct illumination, or with x-rays produced by heating a high-$Z$ cylindrical hohlraum (indirect drive).  Besides its energy-source potential, ICF produces extreme fusion-product fluxes for basic science studies.  Hohlraums are also used to study material properties like opacity.  Hohlraum experiments have been conducted at the National Ignition Facility (NIF) from 2009 to the present \cite{edwards-nic-pop-2013-etal}.  Targets with laser pulses $\gtrsim$10 ns use high density hohlraum gas fills $\gtrsim 0.9$ mg/cm$^3$, typically helium, to tamp high-$Z$ wall expansion.

\begin{figure}
\centering \includegraphics[width=8cm]{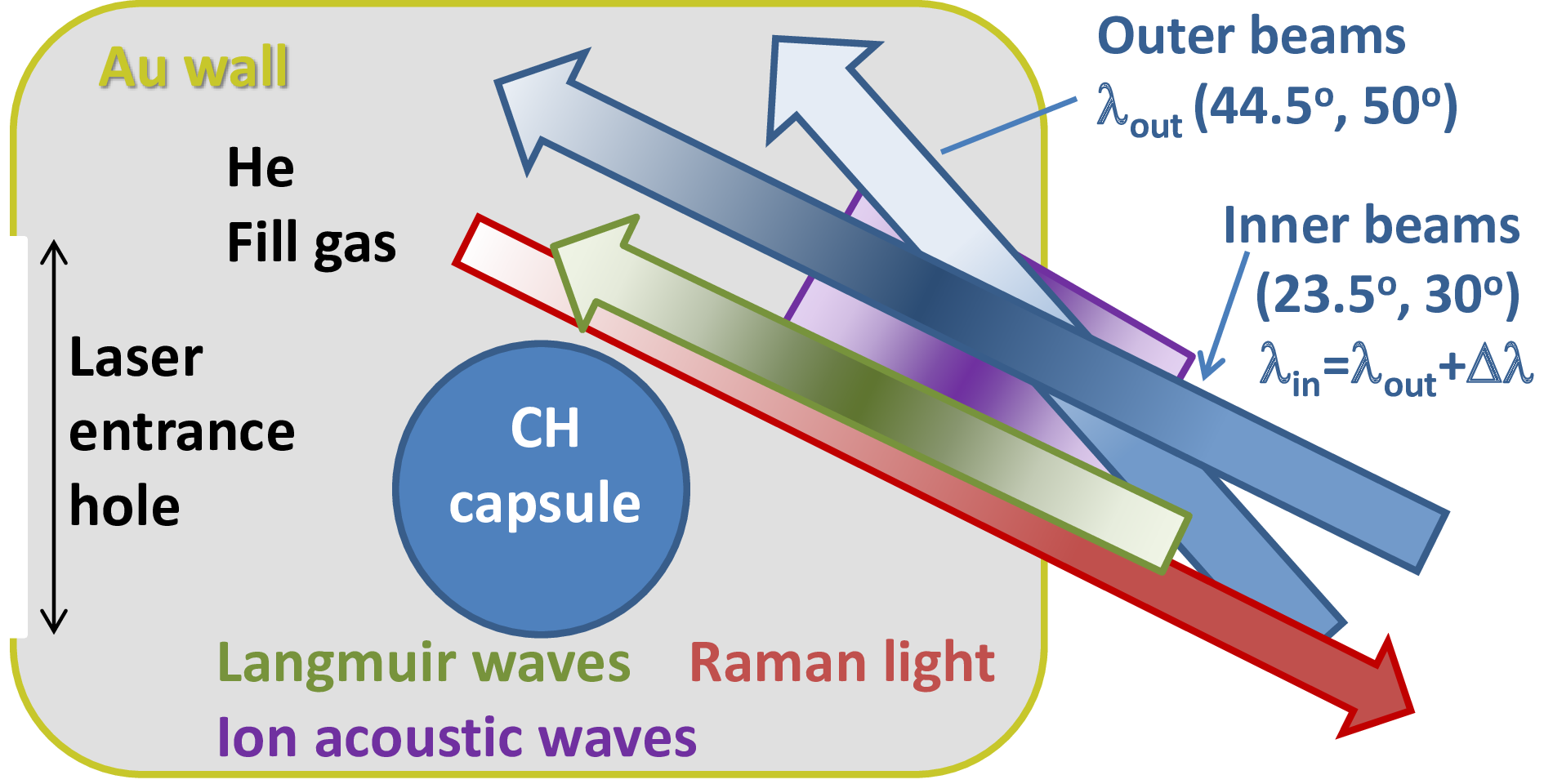}
\caption{\label{fig:cart}Sketch of hohlraum LPI.  Arrows give wave propagation direction, and color darkness indicates intensity.  Outer beams transfer power to inner beams where they overlap.  SRS light from inner beams grows continuously along path, with little absorption.  Langmuir waves are driven by beating of inner-beam laser and SRS light.}
\end{figure}

Laser-plasma interactions are a key aspect of these experiments, as sketched in Fig.\ \ref{fig:cart}. Crossed-beam energy transfer from the outer cones of laser beams (angles to hohlraum axis $\theta=44.5^\circ$ and $50^\circ$) to inner cones ($\theta=23.5^\circ$ and 30$^\circ$) is needed to control implosion symmetry \cite{michel-xbeam-prl-2009}.  CBET is a form of stimulated Brillouin scatter where two light waves beat to drive an ion acoustic wave (IAW) which transfers energy to the light wave with lower frequency in the plasma frame \cite{kruer-cbet-pop-1996}. Shots with high fill density have high inner-beam backward stimulated Raman scatter, or decay of a laser into a scattered light wave and LW. This is detrimental since the scattered light does not produce x-rays, and the LW decays to superthermal electrons which can preheat the fuel and reduce compression.  The LW energy stays in the target, but is spatially redistributed.  This alters symmetry of the x-ray drive and resulting implosion.

LPI processes \footnote{We do not consider two-plasmon decay, which occurs near $n_e/n_{cr}=0.25$ (see Fig.\ref{fig:Tecomp}) and is thus independent of the LPI processes discussed here.} have temporal growth rates (1-10 ps) and spatial gain lengths ($\sim$ speckle length in smoothed beams, $\sim$160 $\mu$m on NIF) much smaller than hydrodynamic scales.  Full LPI modeling therefore requires much more detailed and costly tools than radiation-hydrodynamics codes, such as paraxial-propagation \cite{berger-f3d-pop-1998} or particle-in-cell codes \cite{birdsall-langdon-1989}.  Including LPI effects in rad-hydro codes is challenging: coupling a paraxial and rad-hydro model has been done, but is usually impractical on current computers \cite{glenzer-natphys-2007}.  CBET calculations either post-process plasma conditions from a hydro simulation with no CBET \cite{michel-xbeam-prl-2009, marion-sechel-pop-2016}, or are directly implemented ``inline'' in simulations that describe lasers with ray-tracing \cite{igumenshchev-cbet-pop-2010, marozas-cbet-dpp-2015} or paraxial complex geometric optics \cite{colaitis-cbet-pre-2015}.  SRS is usually treated by removing the escaping light from the incident laser, though recent work has included SRS-produced superthermal electrons in direct-drive hydrodynamic modeling \cite{colaitis-hote-pre-2015}. 

\begin{figure}
\includegraphics[width=9cm]{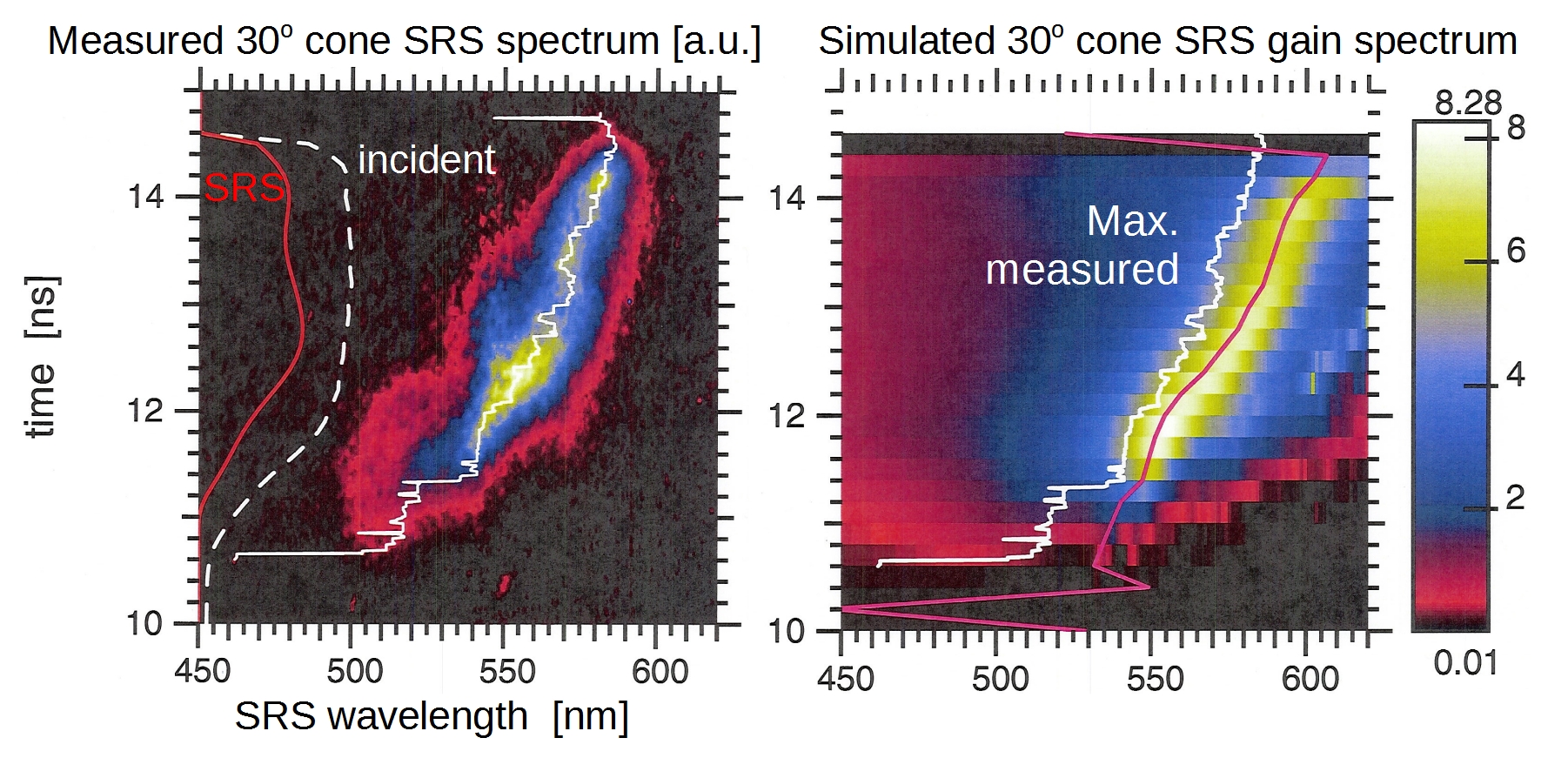}
\caption{\label{fig:fabsgain}Left: measured 30$^\circ$ cone SRS spectrum for NIF shot N121130, with cone incident (white) and escaping SRS (red) powers.  Right: SRS gain exponent spectrum from Lasnex simulation with inline SRS model (magenta: wavelength of maximum gain).}
\end{figure}

In this Letter, we use new, reduced LPI models inline in a rad-hydro code to study the interplay of LPI and hydrodynamics.  We find significant impacts on plasma dynamics and hohlraum irradiation symmetry.  Namely, LPI-driven plasma waves modify plasma conditions and alter CBET in high-fill-density NIF experiments, where CBET can roughly double the inner-beam power, and inner-beam SRS can exceed half the incident power. Unlike prior work, CBET and SRS are modeled together and throughout the target volume, with no assumption about where they occur. We show that LW heating reduces CBET to the inner beams, so that CBET and SRS must be considered jointly.  The SRS light continuously grows as it propagates. LWs are mostly driven just inside the laser entrance hole, which they heat.  Compared to a model where the escaping SRS light is simply removed from the incident laser, the inline model increases the electron temperature where inner and outer beams overlap.  This produces x-rays from the poles rather than equator. 

Our findings help explain several discrepancies between NIF data and hohlraum modeling, namely predictions that almost all outer-beam power is transferred to inner beams.  X-ray images of wall emission shows bright spots corresponding to outer beams, indicating they are not fully depleted \cite{schneider-sxi-rsi-2012-etal}.  Capsule implosion shape data is close to round or oblate (stronger x-ray drive from the poles), but previous modeling gives a strongly prolate shape.

The inline LPI models quantify the processes sketched in Fig.\ \ref{fig:cart}.  We treat light waves as plane waves, and solve steady-state coupled-mode equations for intensities $I_i$ along refracted laser rays.  We model the plasma-wave response with kinetic linear theory in the strong damping limit (advection of plasma waves neglected vs.\ spatial Landau damping).  The SRS model is \textit{post-hoc} in that we specify the escaping SRS powers and wavelengths, which are measured on both inner cones and shown for $\theta=30^\circ$ in Fig.\ \ref{fig:fabsgain}.  This shows the measured SRS wavelength is close to the wavelength of peak SRS gain exponent \cite{strozzi-dep-pop-2008} found from simulated plasma profiles.  The SRS model develops light and Langmuir waves consistent with SRS data, and the resulting spatially-varying energy deposition.

The 192 NIF beams are grouped into 48 quads of four beams with polarization smoothing (two beams linearly polarized orthogonal to the other two).  The inline model for one quad (subscript $X=0$) propagating to $+z$ is
\begin{eqnarray}   
  \p_z I_0 &=& -\kappa_0 I_0 - {g_R\over\omega_R}I_0I_R - \sum_{i=1}^{23}{g_{Ci}\over\omega_i} I_0I_i, \\
 -\p_z I_R &=& -\kappa_R I_R + {g_R\over\omega_0}I_0I_R, \\
 p_L       &=& {\omega_L \over \omega_0\omega_R} g_R I_0I_R, \\
 p_{Ai}   &=& {\omega_{Ai} \over \omega_0\omega_i} g_{Ci} I_0I_i = \alpha_i\delta n_{Ai}^2.
\end{eqnarray}
$\kappa_X$ is the inverse bremsstrahlung absorption rate of wave $X$. $X=i\neq0$ for another quad incident on the same entrance hole, and $X=R$ for quad 0's SRS light wave. For the Langmuir wave ($X=L$),  $\vec k_L=\vec k_0-\vec k_R$, $\omega_L=\omega_0-\omega_R$, $p_L$ is the power deposition density. The IAW for CBET to quad $i$ ($X=Ai$) is analogous, with $L\rightarrow Ai$, $R \rightarrow i$, and $\delta n_{Ai}$ the IAW electron density fluctuation amplitude.  The CBET coupling rate is
\begin{eqnarray}
  g_{Ci} &\equiv& {\pi r_e \over 2m_ec^2}{k_{Ai}^2 \over k_0k_i} \left( 1+\cos^2\theta_i \right) \mathrm{Im}{\chi_e(1+\chi_I) \over 1+\chi_I+\chi_e}, \\
\chi_j &\equiv& -{1 \over 2k_{Ai}^2\lambda_{Dj}^2}Z'\left({\omega_{Ai}-\vec k_{Ai}\cdot\vec u \over k_{Ai}v_{Tj}\sqrt2} \right).
\end{eqnarray}
$k_X=|\vec k_X|$, $ck_X/\omega_X=[1-n_e/n_{cr,X}]^{1/2}$ with $n_{cr,X}$ the critical density for light wave $X$, $\vec u$ is the flow velocity, and $r_e\approx2.82$ fm is the classical electron radius. $\cos\theta_i=\vec k_0\cdot\vec k_i/k_0k_i$, and $(1+\cos^2\theta_i)$ applies for two unpolarized lasers. Ref. \cite{michel-xbet-pop-2009} showed this is appropriate for NIF's polarization scheme.  $\chi_j$ is species $j$ susceptibility, $\chi_I=\sum_{j\in\mathrm{ion}}\chi_j$, $\lambda_{Dj}=(\epsilon_0T_j/n_jZ_j^2e^2)^{1/2}$, $v_{Tj}=(T_j/m_j)^{1/2}$, and $Z$ is the plasma dispersion function.  $g_R$ is obtained from $g_{Ci}$ with $\vec k_{Ai} \rightarrow \vec k_L$, $\omega_{Ai} \rightarrow \omega_L$, $\vec k_i \rightarrow \vec k_R$, $1+\chi_I \rightarrow 1$, and $1+\cos^2\theta \rightarrow 4$ (i.e.\ polarization smoothing does not reduce SRS).

Our model's main assumptions are plane-wave light and linear plasma waves in the strong damping limit.  The first neglects laser speckle structure, which enhances coupling when gain over a speckle length is $\sim1$. For CBET, ref.~\cite{michel-xbet-pop-2009} showed the gain per speckle for any pair of quads is $\ll 1$ in NIF hohlraum conditions, and speckle effects can be neglected.  A post-processing version of our CBET model \cite{michel-crossbeam-pop-2010}, coupled to ``high-flux model'' hydrodynamics described below, has been validated against NIF shape data during the early-time picket \cite{dewald-picket-prl-2013}, as well as peak power when the power or wavelength shifts are not too high \cite{michel-3col-pre-2011}. Nonlinear saturation of CBET-driven IAW's can occur when their amplitudes $\delta n_{Ai}$ exceed the threshold for ion trapping or two-ion-wave decay, both roughly $\delta n_{Ai}/n_e\sim0.01$. We crudely include this by limiting $\delta n_{Ai} = \min[\dnsat, (p_{Ai}/\alpha_i)^{1/2}]$ (the coupling on the right of Eq.\ (1) becomes $\min[I_0I_i, \beta_i(I_0I_i)^{1/2}]$). Our simulations use $\dnsat/n_e=0.01$, though the calculated CBET with $0.004\leq\dnsat/n_e\leq0.1$ is roughly the same.

Our SRS model qualitatively captures results of more advanced, e.g.\ paraxial, simulations \cite{hinkel-aps10-pop-2011}.  These show deep in the hohlraum SRS amplifies thermal noise many orders of magnitude over a few speckle lengths (where SRS may be weakly damped or absolutely unstable).  SRS then grows gradually as it transits the hohlraum, where it is strongly damped and the gain per speckle is small.  Our reduced model applies here, and advances SRS along ray paths in the same direction as the laser.  It stops at the ``seed point'' where SRS becomes convectively stable (absorption exceeds gain), and \textit{calculates} an effective seed power $\Pseed$ which gives the specified escaping SRS power. We find  $\Pseed\sim$ (0.001-0.01)$\times$ incident inner-beam power $P_0$ -- far above thermal noise $\sim10^{-9}P_0$ due to Thomson scatter.  Our model thus captures most SRS power growth and LW heating, since it amplifies $\Pseed$ by ~(50-500)$\times$ to the escaping SRS power.  An improved SRS model may change the seed point or power, but should give similar hydrodynamic effects -- especially since thermal conduction (or superthermal electrons even more so) spreads out the heating. More advanced LPI modeling could help explain the $\Pseed$ we calculate from the measured SRS.

Model (1)--(6) is implemented in the rad-hydro code Lasnex \cite{zimmerman-lasnex-cppcf-1975}, from which we show results. The code describes a laser by rays which carry power instantly $(c \rightarrow \infty)$ along refracted paths.  We present axisymmetric 2D simulations, with intensities found on an auxiliary 3D mesh. We use the ``high-flux model'' for hohlraum simulations \cite{rosen-dca-hedp-2011}, with detailed configuration accounting non-LTE atomic physics \cite{scott-nlte-hedp-2010}, and Spitzer-H{\"a}rm electron heat conduction with flux limit $0.15n_eT_ev_{Te}$.

\begin{figure}
\includegraphics[width=8.7cm]{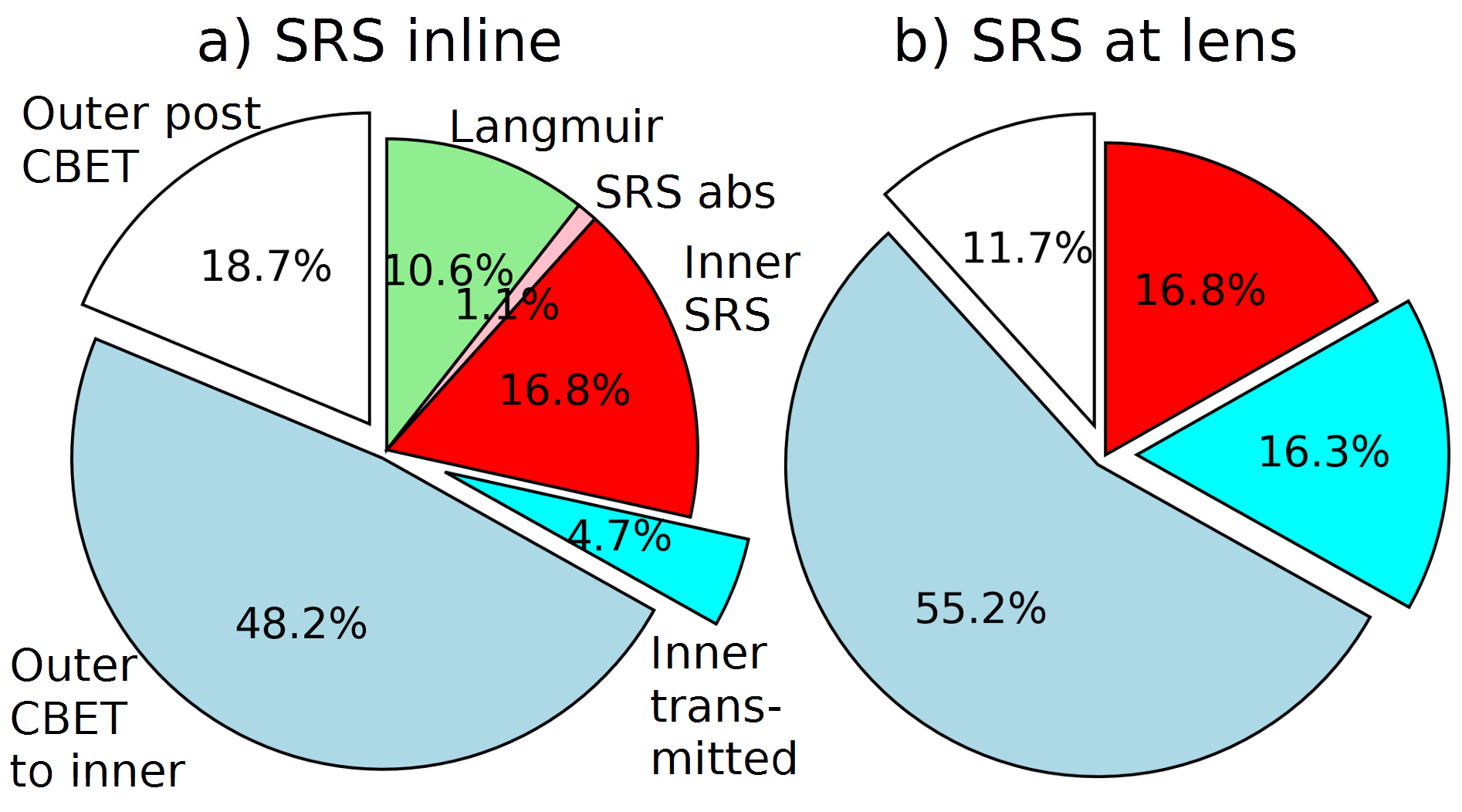}
\caption{\label{fig:pie}Energetics of Lasnex simulations with a) inline SRS model, and b) SRS removed at lens.  ``Outer post CBET:'' incident outer-beam energy not transferred to inners.  ``Inner transmitted'': incident inner energy minus energy to SRS channels.  Inner SRS: escaping inner SRS light.  SRS abs: SRS light absorbed.  Langmuir: energy to LWs.  Energies from 10.5 ns (start of SRS) to 14.8 ns (end of laser pulse), and given as percent of incident laser (1120 kJ).}
\end{figure}

We simulate NIF shot N121130 to show the effects of LPI on hohlraum dynamics. This was an early shot in the high-adiabat campaign \cite{hurricane-hifoot-nat-2014}. 1.27 MJ of frequency-tripled 3$\omega$ ($\lambda=$ 351 nm) laser energy (peak power 350 TW) drove a gold hohlraum filled with 1.45 mg/cm$^3$ of He, and a plastic capsule with D-He3 gas. Cone wavelengths were chosen to give large CBET to the inner (especially $23.5^\circ$) cones: $\lambda_{23}-\lambda_{30}=0.4$ \AA, $\lambda_{30}-\lambda_\mathrm{outer}=2.43$ \AA\ (at $3\omega$). The x-ray emission from the imploded hot spot was moderately oblate, with the amplitude of the $P_2$ Legendre mode -12\% of the $P_0$ mode (average radius), using the contour at 17\% peak brightness (a standard shape measure on NIF). The measured backscatter showed significant inner-cone Raman, low inner-cone Brillouin, and low outer-cone Raman and Brillouin.

\begin{figure}
\includegraphics[width=9cm]{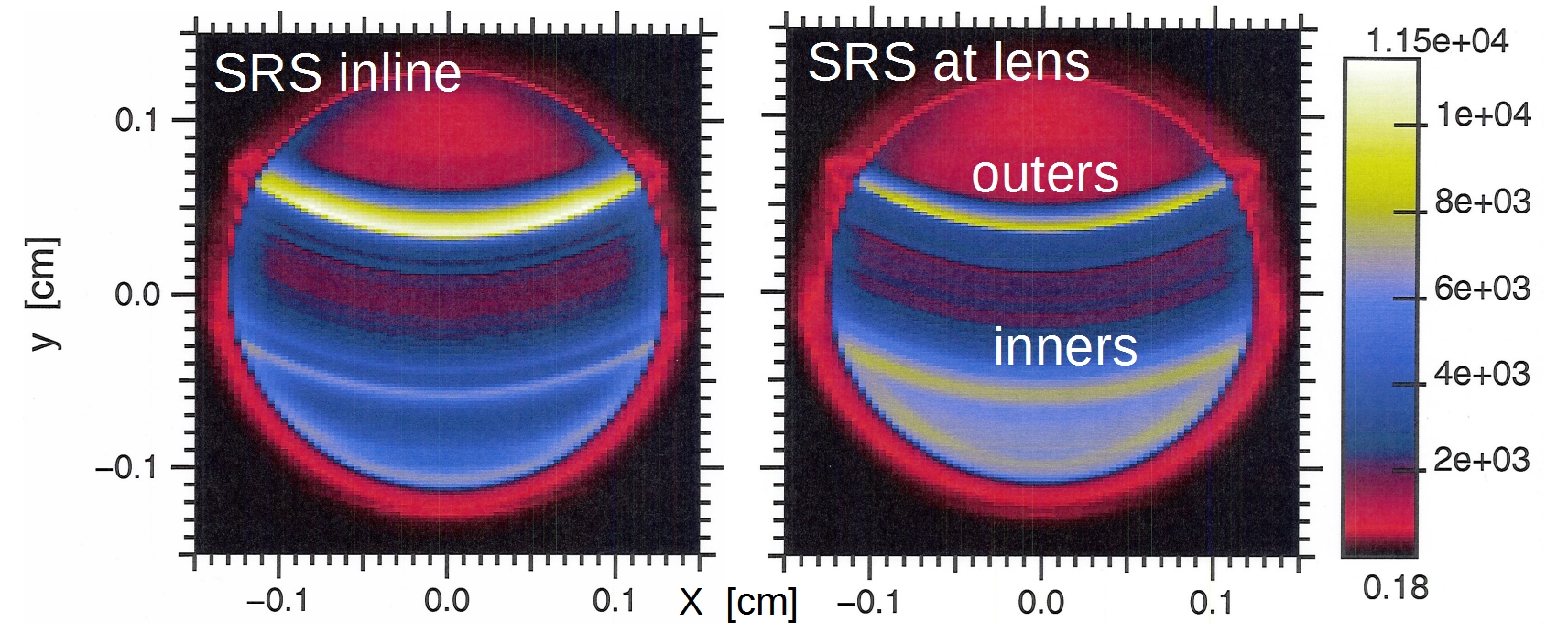}
\caption{\label{fig:sxi}Synthetic x-ray images for Lasnex simulations with inline SRS model (left) and SRS at lens model (right).  Images are symmetrized in azimuth like the simulations. Reduced CBET to inner beams with the inline SRS model gives brighter outer-beam spots. Detector in NIF lower hemisphere 19$^\circ$ to hohlraum axis.  X-ray emission integrated over all time and energies 3 to 5 keV. $y$ is roughly parallel to hohlraum axis.}
\end{figure}

\begin{figure*}
\includegraphics[width=18cm]{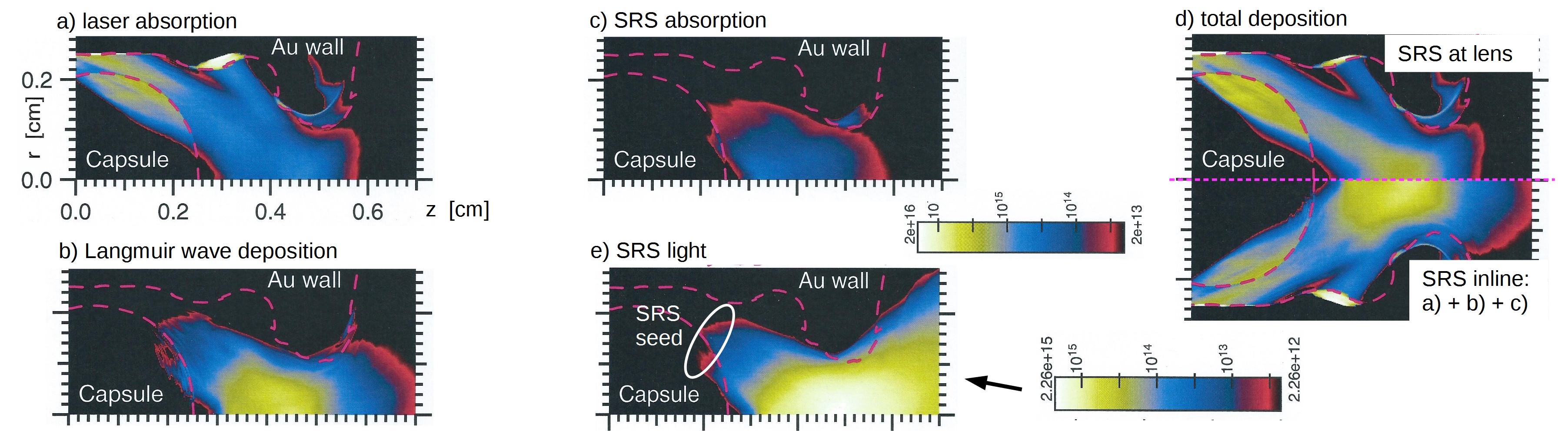}
\caption{\label{fig:srsmaps}Spatial profiles of power densities [W/cm$^3$] from Lasnex simulations at 12.6 ns (time of peak escaping SRS). All panels are from SRS-inline model, except top of d) which is from SRS-at-lens model. a) laser absorption, b) LW deposition, c) SRS light absorption, d) total deposition in SRS-at-lens model (top) and SRS-inline model (bottom). e) SRS light power density, white ellipse indicates seed points where SRS originates.  All panels except e) use same logarithmic colormap. Dashed magenta contours are helium gas boundaries.}
\end{figure*}

To isolate the effects of the inline SRS model, we compare two Lasnex simulations with the inline CBET model.  One uses the inline SRS model.  In the other, the escaping SRS light is removed from the incident laser, with no LW deposition.  This unrealistic ``SRS at lens'' model obtains from Eqs.\ (1)-(6) if $g_R=\delta(\vec x=\vec x_\mathrm{lens})$ and $\omega_R=\omega_0$. The second condition means no energy is deposited to the zero-frequency LW: the same laser energy drives both simulations \footnote{Total x-ray drive in the two simulations is very close: the peak radiation temperature on the capsule is (284.7, 286.7) eV in the (SRS at lens, inline SRS) simulations, and occurs at 14.6 ns in both}.  Figure \ref{fig:pie} gives the energetics. The post-CBET energy on the outers is 60\% higher with the inline-SRS than SRS-at-lens model, while the post-LPI energy on the inners (inner transmitted + outer CBET to inner) is (52.9, 71.5)\% with the (SRS inline, SRS at lens) models. This is reflected in the synthetic image in Fig.\ \ref{fig:sxi} of 3--5 keV x-rays from the entrance hole.  The bright (upper, lower) bands originate from the (outer, inner) beam spots on the hohlraum wall.

Figure \ref{fig:srsmaps} depicts spatial power deposition following the CBET and SRS processes. LW heating is much stronger than SRS absorption, and occurs mostly just inside the entrance hole.  Panel d gives the total heating with the SRS-inline model (sum of panels a, b, and c), and the SRS-at-lens model (just due to laser absorption).  The SRS-inline model has more heating in the entrance hole and outer-beam spots, and less in the inner-beam path.  Panel e shows the SRS power keeps growing until exiting the hohlraum, i.e.\ the SRS gain rate $g_RI_0/\omega_0$ dominates the absorption rate $\kappa_R$.  SRS originates from the ``seed'' region indicated by the circled red color level.

\begin{figure}
\includegraphics[width=7cm]{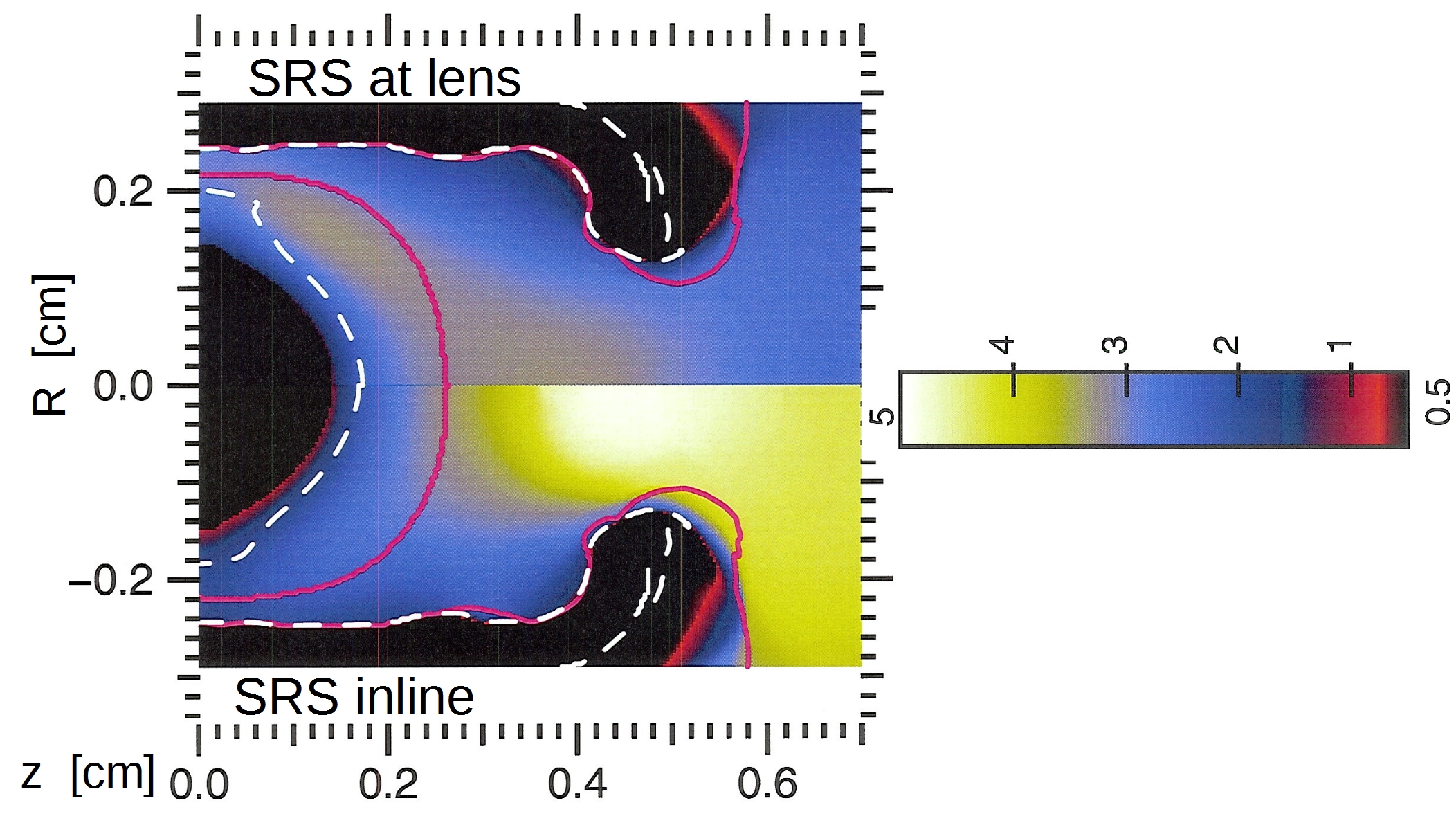}%
\caption{\label{fig:Tecomp} Electron temperature at 12.6 ns from Lasnex simulations shown in Fig.\ \ref{fig:pie}.  $r>0$ has SRS removed at lens, $r<0$ uses inline SRS model with LW deposition and is significantly hotter around the entrance hole. Magenta contours are fill gas boundaries as in Fig.\ \ref{fig:srsmaps}. White dashed contours are $n_e/n_{cr}=0.25$.}
\end{figure}

\begin{figure}
\includegraphics[width=7cm]{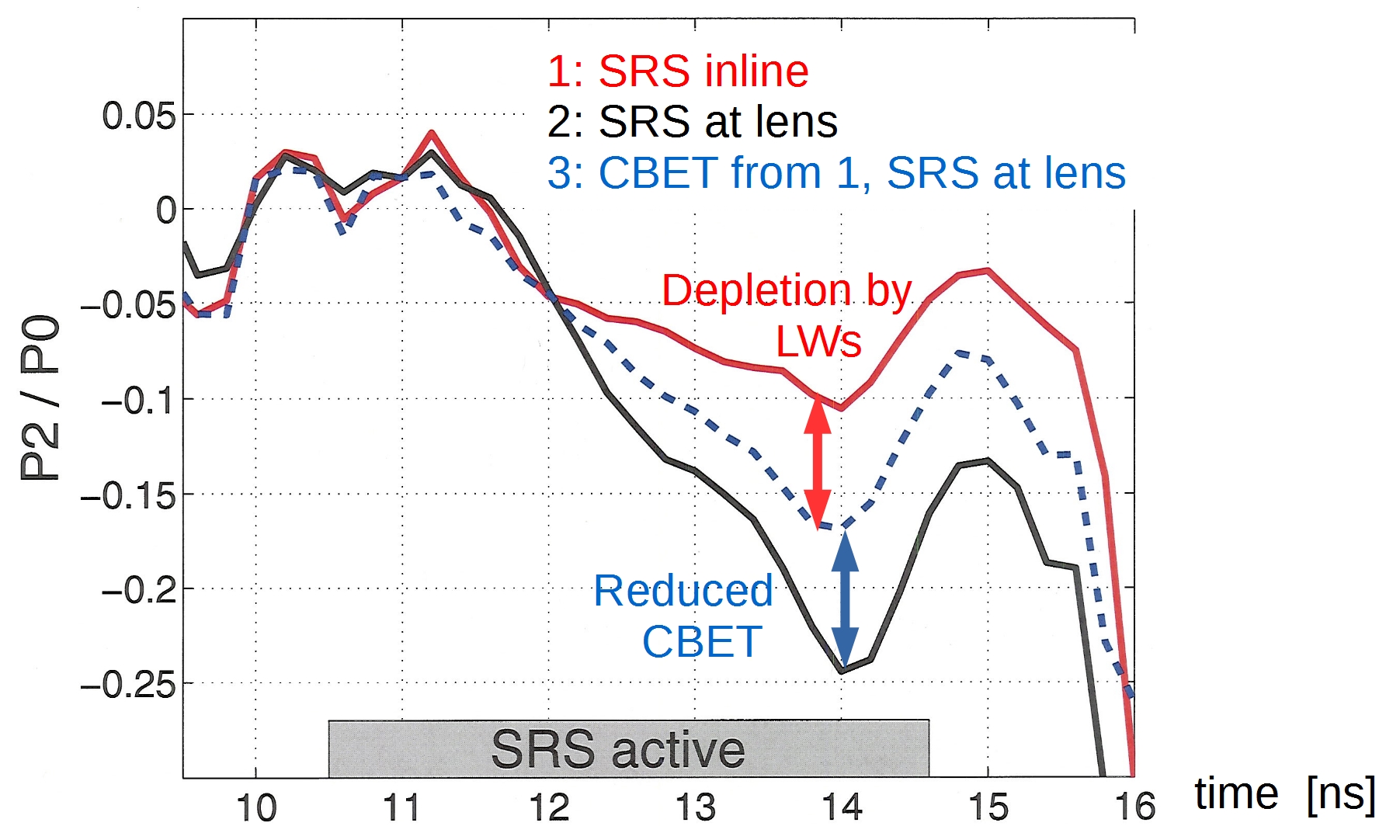}%
\caption{\label{fig:P2}$P_2$ moment of x-ray deposition at ablation front, as fraction of total deposition $P_0$, for x-ray energies 0.5-2 keV. $P_2<0$ for stronger drive from the equator than the pole.  For the inline SRS model, reduced CBET to the inner beams, and depletion by LWs, both reduce the laser intensity at the equator, which makes $P_2$ less negative.}
\end{figure}

The different heating profiles lead to higher entrance hole electron temperature with the inline-SRS model, as shown in Fig.\ \ref{fig:Tecomp}, which reduces CBET to the inners. Recall that in the off-resonant regime $v_{Tj}\gg \omega_{Ai}/k_{Ai}$ appropriate for NIF hohlraums and one ion species, $g_{Ci}\propto (\lambda_i-\lambda_0)Z_jn_eT_j^{1/2}/(T_j+Z_jT_e)^2$ \cite{michel-heating-pop-2013}.  These results can be compared to planned electron-temperature measurements or direct Thomson scatter measurement of LWs in the entrance hole.

The net impact of LPI on symmetry of the x-ray drive is shown in Fig.\ \ref{fig:P2}.  The SRS-inline model (red) gives substantially less equatorial drive than the SRS-at-lens model (black).  A third simulation (blue) separates the effect of reduced CBET, from LW depletion of the inner beams.  We imposed the CBET calculated in the SRS-inline simulation to the incident laser, and removed the escaping SRS from the incident inners.  Comparing the black and blue curves shows the equator drive reduction just due to reduced CBET - the SRS is removed from the incident laser in both cases.  Comparing the blue and red curves isolates the reduction due to LW depletion - the same power is transferred to the inners in both cases.  The two effects are comparable.  LW power is effectively outer-beam power for x-ray symmetry, since they are driven close to the entrance hole.  This is a non-trivial result of the inline SRS model: had the LWs been driven close to the equator wall, they would effectively still be inner-beam power.

To conclude, we have shown a strong effect of laser-plasma interactions on ignition hohlraum plasmas and x-ray drive symmetry.  The Langmuir waves driven by inner-beam Raman scatter are produced near the entrance hole, where they significantly increase the electron temperature.  This reduces energy transfer to the inner beams. Such interplay of hydrodynamics and LPI requires a self-consistent approach, as presented here. The reduced CBET and LW depletion both reduce the inner beam intensity on, and x-ray drive from, the equator wall.  Inline modeling of LPI partially resolves the long-standing over-prediction of equator x-ray drive in NIF hohlraums with high gas fill density.  Future work could improve our reduced model by comparing to more advanced ones with laser speckles and nonlinear kinetics, since direct inclusion of this physics in rad-hydro modeling will be too expensive for the foreseeable future. Accurate laser-driven ICF modeling requires more work on less energetically dominant LPI processes, such as two-plasmon decay and resonance absorption, that can cause unwanted fuel preheat by energetic electrons.  Improved electron transport beyond our simple local model should also be examined.

\begin{acknowledgments}
We thank J.\ A.\ Harte and G.\ B.\ Zimmerman for Lasnex advice.  This work was performed under auspices of the US Dept.\ of Energy by LLNL under Contract DE-AC52-07NA27344.
\end{acknowledgments}


\begin{thebibliography}{29}%
\makeatletter
\providecommand \@ifxundefined [1]{%
 \@ifx{#1\undefined}
}%
\providecommand \@ifnum [1]{%
 \ifnum #1\expandafter \@firstoftwo
 \else \expandafter \@secondoftwo
 \fi
}%
\providecommand \@ifx [1]{%
 \ifx #1\expandafter \@firstoftwo
 \else \expandafter \@secondoftwo
 \fi
}%
\providecommand \natexlab [1]{#1}%
\providecommand \enquote  [1]{``#1''}%
\providecommand \bibnamefont  [1]{#1}%
\providecommand \bibfnamefont [1]{#1}%
\providecommand \citenamefont [1]{#1}%
\providecommand \href@noop [0]{\@secondoftwo}%
\providecommand \href [0]{\begingroup \@sanitize@url \@href}%
\providecommand \@href[1]{\@@startlink{#1}\@@href}%
\providecommand \@@href[1]{\endgroup#1\@@endlink}%
\providecommand \@sanitize@url [0]{\catcode `\\12\catcode `\$12\catcode
  `\&12\catcode `\#12\catcode `\^12\catcode `\_12\catcode `\%12\relax}%
\providecommand \@@startlink[1]{}%
\providecommand \@@endlink[0]{}%
\providecommand \url  [0]{\begingroup\@sanitize@url \@url }%
\providecommand \@url [1]{\endgroup\@href {#1}{\urlprefix }}%
\providecommand \urlprefix  [0]{URL }%
\providecommand \Eprint [0]{\href }%
\providecommand \doibase [0]{http://dx.doi.org/}%
\providecommand \selectlanguage [0]{\@gboble}%
\providecommand \bibinfo  [0]{\@secondoftwo}%
\providecommand \bibfield  [0]{\@secondoftwo}%
\providecommand \translation [1]{[#1]}%
\providecommand \BibitemOpen [0]{}%
\providecommand \bibitemStop [0]{}%
\providecommand \bibitemNoStop [0]{.\EOS\space}%
\providecommand \EOS [0]{\spacefactor3000\relax}%
\providecommand \BibitemShut  [1]{\csname bibitem#1\endcsname}%
\let\auto@bib@innerbib\@empty
\bibitem [{\citenamefont {Fisch}(1987)}]{fisch-currdrive-rmp-1987}%
  \BibitemOpen
  \bibfield  {author} {\bibinfo {author} {\bibfnamefont {N.~J.}\ \bibnamefont
  {Fisch}},\ }\href@noop {} {\bibfield  {journal} {\bibinfo  {journal} {Rev.
  Mod. Phys.}\ }\textbf {\bibinfo {volume} {59}},\ \bibinfo {pages} {175}
  (\bibinfo {year} {1987})}\BibitemShut {NoStop}%
\bibitem [{\citenamefont {Porkolab}(1977)}]{porkolab-parametric-pof-1977}%
  \BibitemOpen
  \bibfield  {author} {\bibinfo {author} {\bibfnamefont {M.}~\bibnamefont
  {Porkolab}},\ }\href@noop {} {\bibfield  {journal} {\bibinfo  {journal}
  {Phys. Fluids}\ }\textbf {\bibinfo {volume} {20}},\ \bibinfo {pages} {2058}
  (\bibinfo {year} {1977})}\BibitemShut {NoStop}%
\bibitem [{\citenamefont {DuBois}\ \emph {et~al.}(2001)\citenamefont {DuBois},
  \citenamefont {Russell}, \citenamefont {Cheung},\ and\ \citenamefont
  {Sulzer}}]{dubois-ionosphere-pop-2001}%
  \BibitemOpen
  \bibfield  {author} {\bibinfo {author} {\bibfnamefont {D.}~\bibnamefont
  {DuBois}}, \bibinfo {author} {\bibfnamefont {D.}~\bibnamefont {Russell}},
  \bibinfo {author} {\bibfnamefont {P.}~\bibnamefont {Cheung}}, \ and\ \bibinfo
  {author} {\bibfnamefont {M.}~\bibnamefont {Sulzer}},\ }\href@noop {}
  {\bibfield  {journal} {\bibinfo  {journal} {Phys. Plasmas}\ }\textbf
  {\bibinfo {volume} {8}},\ \bibinfo {pages} {791} (\bibinfo {year}
  {2001})}\BibitemShut {NoStop}%
\bibitem [{\citenamefont {Malkin}\ \emph {et~al.}(1999)\citenamefont {Malkin},
  \citenamefont {Shvets},\ and\ \citenamefont
  {Fisch}}]{malkin-ramanamp-prl-1999}%
  \BibitemOpen
  \bibfield  {author} {\bibinfo {author} {\bibfnamefont {V.~M.}\ \bibnamefont
  {Malkin}}, \bibinfo {author} {\bibfnamefont {G.}~\bibnamefont {Shvets}}, \
  and\ \bibinfo {author} {\bibfnamefont {N.~J.}\ \bibnamefont {Fisch}},\
  }\href@noop {} {\bibfield  {journal} {\bibinfo  {journal} {Phys. Rev. Lett.}\
  }\textbf {\bibinfo {volume} {82}},\ \bibinfo {pages} {4448} (\bibinfo {year}
  {1999})}\BibitemShut {NoStop}%
\bibitem [{\citenamefont {Edwards}\ \emph {et~al.}(2013)\citenamefont
  {Edwards}, \citenamefont {Patel}, \citenamefont {Lindl}, \citenamefont
  {Atherton}, \citenamefont {Glenzer}, \citenamefont {Haan}, \citenamefont
  {Kilkenny}, \citenamefont {Landen}, \citenamefont {Moses} \emph
  {et~al.}}]{edwards-nic-pop-2013-etal}%
  \BibitemOpen
  \bibfield  {author} {\bibinfo {author} {\bibfnamefont {M.~J.}\ \bibnamefont
  {Edwards}}, \bibinfo {author} {\bibfnamefont {P.~K.}\ \bibnamefont {Patel}},
  \bibinfo {author} {\bibfnamefont {J.~D.}\ \bibnamefont {Lindl}}, \bibinfo
  {author} {\bibfnamefont {L.~J.}\ \bibnamefont {Atherton}}, \bibinfo {author}
  {\bibfnamefont {S.~H.}\ \bibnamefont {Glenzer}}, \bibinfo {author}
  {\bibfnamefont {S.~W.}\ \bibnamefont {Haan}}, \bibinfo {author}
  {\bibfnamefont {J.~D.}\ \bibnamefont {Kilkenny}}, \bibinfo {author}
  {\bibfnamefont {O.~L.}\ \bibnamefont {Landen}}, \bibinfo {author}
  {\bibfnamefont {E.~I.}\ \bibnamefont {Moses}},  \emph {et~al.},\ }\href@noop
  {} {\bibfield  {journal} {\bibinfo  {journal} {Phys. Plasmas}\ }\textbf
  {\bibinfo {volume} {20}},\ \bibinfo {eid} {070501} (\bibinfo {year}
  {2013})}\BibitemShut {NoStop}%
\bibitem [{\citenamefont {Michel}\ \emph
  {et~al.}(2009{\natexlab{a}})\citenamefont {Michel}, \citenamefont {Divol},
  \citenamefont {Williams}, \citenamefont {Weber}, \citenamefont {Thomas},
  \citenamefont {Callahan}, \citenamefont {Haan}, \citenamefont {Salmonson},
  \citenamefont {Dixit} \emph {et~al.}}]{michel-xbeam-prl-2009}%
  \BibitemOpen
  \bibfield  {author} {\bibinfo {author} {\bibfnamefont {P.}~\bibnamefont
  {Michel}}, \bibinfo {author} {\bibfnamefont {L.}~\bibnamefont {Divol}},
  \bibinfo {author} {\bibfnamefont {E.~A.}\ \bibnamefont {Williams}}, \bibinfo
  {author} {\bibfnamefont {S.}~\bibnamefont {Weber}}, \bibinfo {author}
  {\bibfnamefont {C.~A.}\ \bibnamefont {Thomas}}, \bibinfo {author}
  {\bibfnamefont {D.~A.}\ \bibnamefont {Callahan}}, \bibinfo {author}
  {\bibfnamefont {S.~W.}\ \bibnamefont {Haan}}, \bibinfo {author}
  {\bibfnamefont {J.~D.}\ \bibnamefont {Salmonson}}, \bibinfo {author}
  {\bibfnamefont {S.}~\bibnamefont {Dixit}},  \emph {et~al.},\ }\href {\doibase
  10.1103/PhysRevLett.102.025004} {\bibfield  {journal} {\bibinfo  {journal}
  {Phys. Rev. Lett.}\ }\textbf {\bibinfo {volume} {102}},\ \bibinfo {eid}
  {025004} (\bibinfo {year} {2009}{\natexlab{a}})}\BibitemShut {NoStop}%
\bibitem [{\citenamefont {Kruer}\ \emph {et~al.}(1996)\citenamefont {Kruer},
  \citenamefont {Wilks}, \citenamefont {Afeyan},\ and\ \citenamefont
  {Kirkwood}}]{kruer-cbet-pop-1996}%
  \BibitemOpen
  \bibfield  {author} {\bibinfo {author} {\bibfnamefont {W.~L.}\ \bibnamefont
  {Kruer}}, \bibinfo {author} {\bibfnamefont {S.~C.}\ \bibnamefont {Wilks}},
  \bibinfo {author} {\bibfnamefont {B.~B.}\ \bibnamefont {Afeyan}}, \ and\
  \bibinfo {author} {\bibfnamefont {R.~K.}\ \bibnamefont {Kirkwood}},\ }\href
  {\doibase http://dx.doi.org/10.1063/1.871863} {\bibfield  {journal} {\bibinfo
   {journal} {Phys. Plasmas}\ }\textbf {\bibinfo {volume} {3}},\ \bibinfo
  {pages} {382} (\bibinfo {year} {1996})}\BibitemShut {NoStop}%
\bibitem [{Note1()}]{Note1}%
  \BibitemOpen
  \bibinfo {note} {We do not consider two-plasmon decay, which occurs near
  $n_e/n_{cr}=0.25$ (see Fig.\ref {fig:Tecomp}) and is thus independent of the
  LPI processes discussed here.}\BibitemShut {Stop}%
\bibitem [{\citenamefont {Berger}\ \emph {et~al.}(1998)\citenamefont {Berger},
  \citenamefont {Still}, \citenamefont {Williams},\ and\ \citenamefont
  {Langdon}}]{berger-f3d-pop-1998}%
  \BibitemOpen
  \bibfield  {author} {\bibinfo {author} {\bibfnamefont {R.~L.}\ \bibnamefont
  {Berger}}, \bibinfo {author} {\bibfnamefont {C.~H.}\ \bibnamefont {Still}},
  \bibinfo {author} {\bibfnamefont {E.~A.}\ \bibnamefont {Williams}}, \ and\
  \bibinfo {author} {\bibfnamefont {A.~B.}\ \bibnamefont {Langdon}},\
  }\href@noop {} {\bibfield  {journal} {\bibinfo  {journal} {Phys. Plasmas}\
  }\textbf {\bibinfo {volume} {5}},\ \bibinfo {pages} {4337} (\bibinfo {year}
  {1998})}\BibitemShut {NoStop}%
\bibitem [{\citenamefont {Birdsall}\ and\ \citenamefont
  {Langdon}(2005)}]{birdsall-langdon-1989}%
  \BibitemOpen
  \bibfield  {author} {\bibinfo {author} {\bibfnamefont {C.~K.}\ \bibnamefont
  {Birdsall}}\ and\ \bibinfo {author} {\bibfnamefont {A.~B.}\ \bibnamefont
  {Langdon}},\ }\href@noop {} {\emph {\bibinfo {title} {Plasma Physics via
  Computer Simulation}}}\ (\bibinfo  {publisher} {Taylor \& Francis Group},\
  \bibinfo {address} {New York, NY},\ \bibinfo {year} {2005})\BibitemShut
  {NoStop}%
\bibitem [{\citenamefont {Glenzer}\ \emph {et~al.}(2007)\citenamefont
  {Glenzer}, \citenamefont {Froula}, \citenamefont {Divol}, \citenamefont
  {Dorr}, \citenamefont {Berger}, \citenamefont {Dixit}, \citenamefont
  {Hammel}, \citenamefont {Haynam}, \citenamefont {Hittinger}, \citenamefont
  {Holder} \emph {et~al.}}]{glenzer-natphys-2007}%
  \BibitemOpen
  \bibfield  {author} {\bibinfo {author} {\bibfnamefont {S.~H.}\ \bibnamefont
  {Glenzer}}, \bibinfo {author} {\bibfnamefont {D.}~\bibnamefont {Froula}},
  \bibinfo {author} {\bibfnamefont {L.}~\bibnamefont {Divol}}, \bibinfo
  {author} {\bibfnamefont {M.}~\bibnamefont {Dorr}}, \bibinfo {author}
  {\bibfnamefont {R.}~\bibnamefont {Berger}}, \bibinfo {author} {\bibfnamefont
  {S.}~\bibnamefont {Dixit}}, \bibinfo {author} {\bibfnamefont
  {B.}~\bibnamefont {Hammel}}, \bibinfo {author} {\bibfnamefont
  {C.}~\bibnamefont {Haynam}}, \bibinfo {author} {\bibfnamefont
  {J.}~\bibnamefont {Hittinger}}, \bibinfo {author} {\bibfnamefont
  {J.}~\bibnamefont {Holder}},  \emph {et~al.},\ }\href@noop {} {\bibfield
  {journal} {\bibinfo  {journal} {Nature Phys.}\ }\textbf {\bibinfo {volume}
  {3}},\ \bibinfo {pages} {716} (\bibinfo {year} {2007})}\BibitemShut {NoStop}%
\bibitem [{\citenamefont {Marion}\ \emph {et~al.}(2016)\citenamefont {Marion},
  \citenamefont {Debayle}, \citenamefont {Masson-Laborde}, \citenamefont
  {Loiseau},\ and\ \citenamefont {Casanova}}]{marion-sechel-pop-2016}%
  \BibitemOpen
  \bibfield  {author} {\bibinfo {author} {\bibfnamefont {D.}~\bibnamefont
  {Marion}}, \bibinfo {author} {\bibfnamefont {A.}~\bibnamefont {Debayle}},
  \bibinfo {author} {\bibfnamefont {P.-E.}\ \bibnamefont {Masson-Laborde}},
  \bibinfo {author} {\bibfnamefont {P.}~\bibnamefont {Loiseau}}, \ and\
  \bibinfo {author} {\bibfnamefont {M.}~\bibnamefont {Casanova}},\ }\href@noop
  {} {\bibfield  {journal} {\bibinfo  {journal} {Phys. Plasmas}\ }\textbf
  {\bibinfo {volume} {23}},\ \bibinfo {pages} {052705} (\bibinfo {year}
  {2016})}\BibitemShut {NoStop}%
\bibitem [{\citenamefont {Igumenshchev}\ \emph {et~al.}(2010)\citenamefont
  {Igumenshchev}, \citenamefont {Edgell}, \citenamefont {Goncharov},
  \citenamefont {Delettrez}, \citenamefont {Maximov}, \citenamefont {Myatt},
  \citenamefont {Seka}, \citenamefont {Shvydky}, \citenamefont {Skupsky},\ and\
  \citenamefont {Stoeckl}}]{igumenshchev-cbet-pop-2010}%
  \BibitemOpen
  \bibfield  {author} {\bibinfo {author} {\bibfnamefont {I.}~\bibnamefont
  {Igumenshchev}}, \bibinfo {author} {\bibfnamefont {D.}~\bibnamefont
  {Edgell}}, \bibinfo {author} {\bibfnamefont {V.}~\bibnamefont {Goncharov}},
  \bibinfo {author} {\bibfnamefont {J.}~\bibnamefont {Delettrez}}, \bibinfo
  {author} {\bibfnamefont {A.}~\bibnamefont {Maximov}}, \bibinfo {author}
  {\bibfnamefont {J.}~\bibnamefont {Myatt}}, \bibinfo {author} {\bibfnamefont
  {W.}~\bibnamefont {Seka}}, \bibinfo {author} {\bibfnamefont {A.}~\bibnamefont
  {Shvydky}}, \bibinfo {author} {\bibfnamefont {S.}~\bibnamefont {Skupsky}}, \
  and\ \bibinfo {author} {\bibfnamefont {C.}~\bibnamefont {Stoeckl}},\
  }\href@noop {} {\bibfield  {journal} {\bibinfo  {journal} {Phys. Plasmas}\
  }\textbf {\bibinfo {volume} {17}},\ \bibinfo {pages} {122708} (\bibinfo
  {year} {2010})}\BibitemShut {NoStop}%
\bibitem [{\citenamefont {Marozas}\ \emph {et~al.}(2015)\citenamefont
  {Marozas}, \citenamefont {Collins}, \citenamefont {McKenty},\ and\
  \citenamefont {Zuegel}}]{marozas-cbet-dpp-2015}%
  \BibitemOpen
  \bibfield  {author} {\bibinfo {author} {\bibfnamefont {J.}~\bibnamefont
  {Marozas}}, \bibinfo {author} {\bibfnamefont {T.}~\bibnamefont {Collins}},
  \bibinfo {author} {\bibfnamefont {P.}~\bibnamefont {McKenty}}, \ and\
  \bibinfo {author} {\bibfnamefont {J.}~\bibnamefont {Zuegel}},\ }\href@noop {}
  {\bibfield  {journal} {\bibinfo  {journal} {Bull. Am. Phys. Soc.}\ }\textbf
  {\bibinfo {volume} {60}} (\bibinfo {year} {2015})}\BibitemShut {NoStop}%
\bibitem [{\citenamefont {Cola{\"\i}tis}\ \emph
  {et~al.}(2015{\natexlab{a}})\citenamefont {Cola{\"\i}tis}, \citenamefont
  {Duchateau}, \citenamefont {Ribeyre},\ and\ \citenamefont
  {Tikhonchuk}}]{colaitis-cbet-pre-2015}%
  \BibitemOpen
  \bibfield  {author} {\bibinfo {author} {\bibfnamefont {A.}~\bibnamefont
  {Cola{\"\i}tis}}, \bibinfo {author} {\bibfnamefont {G.}~\bibnamefont
  {Duchateau}}, \bibinfo {author} {\bibfnamefont {X.}~\bibnamefont {Ribeyre}},
  \ and\ \bibinfo {author} {\bibfnamefont {V.}~\bibnamefont {Tikhonchuk}},\
  }\href@noop {} {\bibfield  {journal} {\bibinfo  {journal} {Phys. Rev. E}\
  }\textbf {\bibinfo {volume} {91}},\ \bibinfo {pages} {013102} (\bibinfo
  {year} {2015}{\natexlab{a}})}\BibitemShut {NoStop}%
\bibitem [{\citenamefont {Cola{\"\i}tis}\ \emph
  {et~al.}(2015{\natexlab{b}})\citenamefont {Cola{\"\i}tis}, \citenamefont
  {Duchateau}, \citenamefont {Ribeyre}, \citenamefont {Maheut}, \citenamefont
  {Boutoux}, \citenamefont {Antonelli}, \citenamefont {Nicola{\"\i}},
  \citenamefont {Batani},\ and\ \citenamefont
  {Tikhonchuk}}]{colaitis-hote-pre-2015}%
  \BibitemOpen
  \bibfield  {author} {\bibinfo {author} {\bibfnamefont {A.}~\bibnamefont
  {Cola{\"\i}tis}}, \bibinfo {author} {\bibfnamefont {G.}~\bibnamefont
  {Duchateau}}, \bibinfo {author} {\bibfnamefont {X.}~\bibnamefont {Ribeyre}},
  \bibinfo {author} {\bibfnamefont {Y.}~\bibnamefont {Maheut}}, \bibinfo
  {author} {\bibfnamefont {G.}~\bibnamefont {Boutoux}}, \bibinfo {author}
  {\bibfnamefont {L.}~\bibnamefont {Antonelli}}, \bibinfo {author}
  {\bibfnamefont {P.}~\bibnamefont {Nicola{\"\i}}}, \bibinfo {author}
  {\bibfnamefont {D.}~\bibnamefont {Batani}}, \ and\ \bibinfo {author}
  {\bibfnamefont {V.}~\bibnamefont {Tikhonchuk}},\ }\href@noop {} {\bibfield
  {journal} {\bibinfo  {journal} {Phys. Rev. E}\ }\textbf {\bibinfo {volume}
  {92}},\ \bibinfo {pages} {041101} (\bibinfo {year}
  {2015}{\natexlab{b}})}\BibitemShut {NoStop}%
\bibitem [{\citenamefont {Schneider}\ \emph {et~al.}(2012)\citenamefont
  {Schneider}, \citenamefont {Meezan}, \citenamefont {Alvarez}, \citenamefont
  {Alameda}, \citenamefont {Baker}, \citenamefont {Bell}, \citenamefont
  {Bradley}, \citenamefont {Callahan}, \citenamefont {Celeste} \emph
  {et~al.}}]{schneider-sxi-rsi-2012-etal}%
  \BibitemOpen
  \bibfield  {author} {\bibinfo {author} {\bibfnamefont {M.~B.}\ \bibnamefont
  {Schneider}}, \bibinfo {author} {\bibfnamefont {N.~B.}\ \bibnamefont
  {Meezan}}, \bibinfo {author} {\bibfnamefont {S.~S.}\ \bibnamefont {Alvarez}},
  \bibinfo {author} {\bibfnamefont {J.}~\bibnamefont {Alameda}}, \bibinfo
  {author} {\bibfnamefont {S.}~\bibnamefont {Baker}}, \bibinfo {author}
  {\bibfnamefont {P.~M.}\ \bibnamefont {Bell}}, \bibinfo {author}
  {\bibfnamefont {D.~K.}\ \bibnamefont {Bradley}}, \bibinfo {author}
  {\bibfnamefont {D.~A.}\ \bibnamefont {Callahan}}, \bibinfo {author}
  {\bibfnamefont {J.~R.}\ \bibnamefont {Celeste}},  \emph {et~al.},\ }\href
  {http://scitation.aip.org/content/aip/journal/rsi/83/10/10.1063/1.4732850}
  {\bibfield  {journal} {\bibinfo  {journal} {Rev. Sci. Instrum.}\ }\textbf
  {\bibinfo {volume} {83}},\ \bibinfo {eid} {10E525} (\bibinfo {year}
  {2012})}\BibitemShut {NoStop}%
\bibitem [{\citenamefont {Strozzi}\ \emph {et~al.}(2008)\citenamefont
  {Strozzi}, \citenamefont {Williams}, \citenamefont {Hinkel}, \citenamefont
  {Froula}, \citenamefont {London},\ and\ \citenamefont
  {Callahan}}]{strozzi-dep-pop-2008}%
  \BibitemOpen
  \bibfield  {author} {\bibinfo {author} {\bibfnamefont {D.~J.}\ \bibnamefont
  {Strozzi}}, \bibinfo {author} {\bibfnamefont {E.~A.}\ \bibnamefont
  {Williams}}, \bibinfo {author} {\bibfnamefont {D.~E.}\ \bibnamefont
  {Hinkel}}, \bibinfo {author} {\bibfnamefont {D.~H.}\ \bibnamefont {Froula}},
  \bibinfo {author} {\bibfnamefont {R.~A.}\ \bibnamefont {London}}, \ and\
  \bibinfo {author} {\bibfnamefont {D.~A.}\ \bibnamefont {Callahan}},\ }\href
  {\doibase 10.1063/1.2992522} {\bibfield  {journal} {\bibinfo  {journal}
  {Phys. Plasmas}\ }\textbf {\bibinfo {volume} {15}},\ \bibinfo {eid} {102703}
  (\bibinfo {year} {2008})}\BibitemShut {NoStop}%
\bibitem [{\citenamefont {Michel}\ \emph
  {et~al.}(2009{\natexlab{b}})\citenamefont {Michel}, \citenamefont {Divol},
  \citenamefont {Williams}, \citenamefont {Thomas}, \citenamefont {Callahan},
  \citenamefont {Weber}, \citenamefont {Haan}, \citenamefont {Salmonson},
  \citenamefont {Meezan}, \citenamefont {Landen} \emph
  {et~al.}}]{michel-xbet-pop-2009}%
  \BibitemOpen
  \bibfield  {author} {\bibinfo {author} {\bibfnamefont {P.}~\bibnamefont
  {Michel}}, \bibinfo {author} {\bibfnamefont {L.}~\bibnamefont {Divol}},
  \bibinfo {author} {\bibfnamefont {E.}~\bibnamefont {Williams}}, \bibinfo
  {author} {\bibfnamefont {C.}~\bibnamefont {Thomas}}, \bibinfo {author}
  {\bibfnamefont {D.}~\bibnamefont {Callahan}}, \bibinfo {author}
  {\bibfnamefont {S.}~\bibnamefont {Weber}}, \bibinfo {author} {\bibfnamefont
  {S.}~\bibnamefont {Haan}}, \bibinfo {author} {\bibfnamefont {J.}~\bibnamefont
  {Salmonson}}, \bibinfo {author} {\bibfnamefont {N.}~\bibnamefont {Meezan}},
  \bibinfo {author} {\bibfnamefont {O.}~\bibnamefont {Landen}},  \emph
  {et~al.},\ }\href@noop {} {\bibfield  {journal} {\bibinfo  {journal} {Phys.
  Plasmas}\ }\textbf {\bibinfo {volume} {16}},\ \bibinfo {pages} {042702}
  (\bibinfo {year} {2009}{\natexlab{b}})}\BibitemShut {NoStop}%
\bibitem [{\citenamefont {Michel}\ \emph {et~al.}(2010)\citenamefont {Michel},
  \citenamefont {Glenzer}, \citenamefont {Divol}, \citenamefont {Bradley},
  \citenamefont {Callahan}, \citenamefont {Dixit}, \citenamefont {Glenn},
  \citenamefont {Hinkel}, \citenamefont {Kirkwood} \emph
  {et~al.}}]{michel-crossbeam-pop-2010}%
  \BibitemOpen
  \bibfield  {author} {\bibinfo {author} {\bibfnamefont {P.}~\bibnamefont
  {Michel}}, \bibinfo {author} {\bibfnamefont {S.~H.}\ \bibnamefont {Glenzer}},
  \bibinfo {author} {\bibfnamefont {L.}~\bibnamefont {Divol}}, \bibinfo
  {author} {\bibfnamefont {D.~K.}\ \bibnamefont {Bradley}}, \bibinfo {author}
  {\bibfnamefont {D.}~\bibnamefont {Callahan}}, \bibinfo {author}
  {\bibfnamefont {S.}~\bibnamefont {Dixit}}, \bibinfo {author} {\bibfnamefont
  {S.}~\bibnamefont {Glenn}}, \bibinfo {author} {\bibfnamefont
  {D.}~\bibnamefont {Hinkel}}, \bibinfo {author} {\bibfnamefont {R.~K.}\
  \bibnamefont {Kirkwood}},  \emph {et~al.},\ }\href {\doibase
  10.1063/1.3325733} {\bibfield  {journal} {\bibinfo  {journal} {Phys.
  Plasmas}\ }\textbf {\bibinfo {volume} {17}},\ \bibinfo {eid} {056305}
  (\bibinfo {year} {2010})}\BibitemShut {NoStop}%
\bibitem [{\citenamefont {Dewald}\ \emph {et~al.}(2013)\citenamefont {Dewald},
  \citenamefont {Milovich}, \citenamefont {Michel}, \citenamefont {Landen},
  \citenamefont {Kline}, \citenamefont {Glenn}, \citenamefont {Jones},
  \citenamefont {Kalantar}, \citenamefont {Pak}, \citenamefont {Robey} \emph
  {et~al.}}]{dewald-picket-prl-2013}%
  \BibitemOpen
  \bibfield  {author} {\bibinfo {author} {\bibfnamefont {E.}~\bibnamefont
  {Dewald}}, \bibinfo {author} {\bibfnamefont {J.}~\bibnamefont {Milovich}},
  \bibinfo {author} {\bibfnamefont {P.}~\bibnamefont {Michel}}, \bibinfo
  {author} {\bibfnamefont {O.}~\bibnamefont {Landen}}, \bibinfo {author}
  {\bibfnamefont {J.}~\bibnamefont {Kline}}, \bibinfo {author} {\bibfnamefont
  {S.}~\bibnamefont {Glenn}}, \bibinfo {author} {\bibfnamefont
  {O.}~\bibnamefont {Jones}}, \bibinfo {author} {\bibfnamefont
  {D.}~\bibnamefont {Kalantar}}, \bibinfo {author} {\bibfnamefont
  {A.}~\bibnamefont {Pak}}, \bibinfo {author} {\bibfnamefont {H.}~\bibnamefont
  {Robey}},  \emph {et~al.},\ }\href@noop {} {\bibfield  {journal} {\bibinfo
  {journal} {Phys. Rev. Lett.}\ }\textbf {\bibinfo {volume} {111}},\ \bibinfo
  {pages} {235001} (\bibinfo {year} {2013})}\BibitemShut {NoStop}%
\bibitem [{\citenamefont {Michel}\ \emph {et~al.}(2011)\citenamefont {Michel},
  \citenamefont {Divol}, \citenamefont {Town}, \citenamefont {Rosen},
  \citenamefont {Callahan}, \citenamefont {Meezan}, \citenamefont {Schneider},
  \citenamefont {Kyrala}, \citenamefont {Moody} \emph
  {et~al.}}]{michel-3col-pre-2011}%
  \BibitemOpen
  \bibfield  {author} {\bibinfo {author} {\bibfnamefont {P.}~\bibnamefont
  {Michel}}, \bibinfo {author} {\bibfnamefont {L.}~\bibnamefont {Divol}},
  \bibinfo {author} {\bibfnamefont {R.~P.~J.}\ \bibnamefont {Town}}, \bibinfo
  {author} {\bibfnamefont {M.~D.}\ \bibnamefont {Rosen}}, \bibinfo {author}
  {\bibfnamefont {D.~A.}\ \bibnamefont {Callahan}}, \bibinfo {author}
  {\bibfnamefont {N.~B.}\ \bibnamefont {Meezan}}, \bibinfo {author}
  {\bibfnamefont {M.~B.}\ \bibnamefont {Schneider}}, \bibinfo {author}
  {\bibfnamefont {G.~A.}\ \bibnamefont {Kyrala}}, \bibinfo {author}
  {\bibfnamefont {J.~D.}\ \bibnamefont {Moody}},  \emph {et~al.},\ }\href
  {\doibase 10.1103/PhysRevE.83.046409} {\bibfield  {journal} {\bibinfo
  {journal} {Phys. Rev. E}\ }\textbf {\bibinfo {volume} {83}},\ \bibinfo
  {pages} {046409} (\bibinfo {year} {2011})}\BibitemShut {NoStop}%
\bibitem [{\citenamefont {Hinkel}\ \emph {et~al.}(2011)\citenamefont {Hinkel},
  \citenamefont {Rosen}, \citenamefont {Williams}, \citenamefont {Langdon},
  \citenamefont {Still}, \citenamefont {Callahan}, \citenamefont {Moody},
  \citenamefont {Michel}, \citenamefont {Town}, \citenamefont {London},\ and\
  \citenamefont {Langer}}]{hinkel-aps10-pop-2011}%
  \BibitemOpen
  \bibfield  {author} {\bibinfo {author} {\bibfnamefont {D.~E.}\ \bibnamefont
  {Hinkel}}, \bibinfo {author} {\bibfnamefont {M.~D.}\ \bibnamefont {Rosen}},
  \bibinfo {author} {\bibfnamefont {E.~A.}\ \bibnamefont {Williams}}, \bibinfo
  {author} {\bibfnamefont {A.~B.}\ \bibnamefont {Langdon}}, \bibinfo {author}
  {\bibfnamefont {C.~H.}\ \bibnamefont {Still}}, \bibinfo {author}
  {\bibfnamefont {D.~A.}\ \bibnamefont {Callahan}}, \bibinfo {author}
  {\bibfnamefont {J.~D.}\ \bibnamefont {Moody}}, \bibinfo {author}
  {\bibfnamefont {P.~A.}\ \bibnamefont {Michel}}, \bibinfo {author}
  {\bibfnamefont {R.~P.~J.}\ \bibnamefont {Town}}, \bibinfo {author}
  {\bibfnamefont {R.~A.}\ \bibnamefont {London}}, \ and\ \bibinfo {author}
  {\bibfnamefont {S.~H.}\ \bibnamefont {Langer}},\ }\href {\doibase
  10.1063/1.3577836} {\bibfield  {journal} {\bibinfo  {journal} {Phys.
  Plasmas}\ }\textbf {\bibinfo {volume} {18}},\ \bibinfo {eid} {056312}
  (\bibinfo {year} {2011})}\BibitemShut {NoStop}%
\bibitem [{\citenamefont {Zimmerman}\ and\ \citenamefont
  {Kruer}(1975)}]{zimmerman-lasnex-cppcf-1975}%
  \BibitemOpen
  \bibfield  {author} {\bibinfo {author} {\bibfnamefont {G.}~\bibnamefont
  {Zimmerman}}\ and\ \bibinfo {author} {\bibfnamefont {W.~L.}\ \bibnamefont
  {Kruer}},\ }\href@noop {} {\bibfield  {journal} {\bibinfo  {journal}
  {Comments Plasma Phys. Controlled Fusion}\ }\textbf {\bibinfo {volume} {2}},\
  \bibinfo {pages} {85} (\bibinfo {year} {1975})}\BibitemShut {NoStop}%
\bibitem [{\citenamefont {Rosen}\ \emph {et~al.}(2011)\citenamefont {Rosen},
  \citenamefont {Scott}, \citenamefont {Hinkel}, \citenamefont {Williams},
  \citenamefont {Callahan}, \citenamefont {Town}, \citenamefont {Divol},
  \citenamefont {Michel}, \citenamefont {Kruer}, \citenamefont {Suter},
  \citenamefont {London}, \citenamefont {Harte},\ and\ \citenamefont
  {Zimmerman}}]{rosen-dca-hedp-2011}%
  \BibitemOpen
  \bibfield  {author} {\bibinfo {author} {\bibfnamefont {M.}~\bibnamefont
  {Rosen}}, \bibinfo {author} {\bibfnamefont {H.}~\bibnamefont {Scott}},
  \bibinfo {author} {\bibfnamefont {D.}~\bibnamefont {Hinkel}}, \bibinfo
  {author} {\bibfnamefont {E.}~\bibnamefont {Williams}}, \bibinfo {author}
  {\bibfnamefont {D.}~\bibnamefont {Callahan}}, \bibinfo {author}
  {\bibfnamefont {R.}~\bibnamefont {Town}}, \bibinfo {author} {\bibfnamefont
  {L.}~\bibnamefont {Divol}}, \bibinfo {author} {\bibfnamefont
  {P.}~\bibnamefont {Michel}}, \bibinfo {author} {\bibfnamefont
  {W.}~\bibnamefont {Kruer}}, \bibinfo {author} {\bibfnamefont
  {L.}~\bibnamefont {Suter}}, \bibinfo {author} {\bibfnamefont
  {R.}~\bibnamefont {London}}, \bibinfo {author} {\bibfnamefont
  {J.}~\bibnamefont {Harte}}, \ and\ \bibinfo {author} {\bibfnamefont
  {G.}~\bibnamefont {Zimmerman}},\ }\href {\doibase
  http://dx.doi.org/10.1016/j.hedp.2011.03.008} {\bibfield  {journal} {\bibinfo
   {journal} {High Energy Density Phys.}\ }\textbf {\bibinfo {volume} {7}},\
  \bibinfo {pages} {180 } (\bibinfo {year} {2011})}\BibitemShut {NoStop}%
\bibitem [{\citenamefont {Scott}\ and\ \citenamefont
  {Hansen}(2010)}]{scott-nlte-hedp-2010}%
  \BibitemOpen
  \bibfield  {author} {\bibinfo {author} {\bibfnamefont {H.}~\bibnamefont
  {Scott}}\ and\ \bibinfo {author} {\bibfnamefont {S.}~\bibnamefont {Hansen}},\
  }\href {\doibase 10.1016/j.hedp.2009.07.003} {\bibfield  {journal} {\bibinfo
  {journal} {High Energy Density Phys.}\ }\textbf {\bibinfo {volume} {6}},\
  \bibinfo {pages} {39 } (\bibinfo {year} {2010})}\BibitemShut {NoStop}%
\bibitem [{\citenamefont {Hurricane}\ \emph {et~al.}(2014)\citenamefont
  {Hurricane}, \citenamefont {Callahan}, \citenamefont {Casey}, \citenamefont
  {Celliers}, \citenamefont {Cerjan}, \citenamefont {Dewald}, \citenamefont
  {Dittrich}, \citenamefont {Doppner}, \citenamefont {Hinkel}, \citenamefont
  {Hopkins}, \citenamefont {Kline}, \citenamefont {Le~Pape}, \citenamefont
  {Ma}, \citenamefont {MacPhee}, \citenamefont {Milovich}, \citenamefont {Pak},
  \citenamefont {Park}, \citenamefont {Patel}, \citenamefont {Remington},
  \citenamefont {Salmonson}, \citenamefont {Springer},\ and\ \citenamefont
  {Tommasini}}]{hurricane-hifoot-nat-2014}%
  \BibitemOpen
  \bibfield  {author} {\bibinfo {author} {\bibfnamefont {O.~A.}\ \bibnamefont
  {Hurricane}}, \bibinfo {author} {\bibfnamefont {D.~A.}\ \bibnamefont
  {Callahan}}, \bibinfo {author} {\bibfnamefont {D.~T.}\ \bibnamefont {Casey}},
  \bibinfo {author} {\bibfnamefont {P.~M.}\ \bibnamefont {Celliers}}, \bibinfo
  {author} {\bibfnamefont {C.}~\bibnamefont {Cerjan}}, \bibinfo {author}
  {\bibfnamefont {E.~L.}\ \bibnamefont {Dewald}}, \bibinfo {author}
  {\bibfnamefont {T.~R.}\ \bibnamefont {Dittrich}}, \bibinfo {author}
  {\bibfnamefont {T.}~\bibnamefont {Doppner}}, \bibinfo {author} {\bibfnamefont
  {D.~E.}\ \bibnamefont {Hinkel}}, \bibinfo {author} {\bibfnamefont {L.~F.~B.}\
  \bibnamefont {Hopkins}}, \bibinfo {author} {\bibfnamefont {J.~L.}\
  \bibnamefont {Kline}}, \bibinfo {author} {\bibfnamefont {S.}~\bibnamefont
  {Le~Pape}}, \bibinfo {author} {\bibfnamefont {T.}~\bibnamefont {Ma}},
  \bibinfo {author} {\bibfnamefont {A.~G.}\ \bibnamefont {MacPhee}}, \bibinfo
  {author} {\bibfnamefont {J.~L.}\ \bibnamefont {Milovich}}, \bibinfo {author}
  {\bibfnamefont {A.}~\bibnamefont {Pak}}, \bibinfo {author} {\bibfnamefont
  {H.-S.}\ \bibnamefont {Park}}, \bibinfo {author} {\bibfnamefont {P.~K.}\
  \bibnamefont {Patel}}, \bibinfo {author} {\bibfnamefont {B.~A.}\ \bibnamefont
  {Remington}}, \bibinfo {author} {\bibfnamefont {J.~D.}\ \bibnamefont
  {Salmonson}}, \bibinfo {author} {\bibfnamefont {P.~T.}\ \bibnamefont
  {Springer}}, \ and\ \bibinfo {author} {\bibfnamefont {R.}~\bibnamefont
  {Tommasini}},\ }\href@noop {} {\bibfield  {journal} {\bibinfo  {journal}
  {Nature}\ }\textbf {\bibinfo {volume} {506}},\ \bibinfo {pages} {343}
  (\bibinfo {year} {2014})}\BibitemShut {NoStop}%
\bibitem [{Note2()}]{Note2}%
  \BibitemOpen
  \bibinfo {note} {Total x-ray drive in the two simulations is very close: the
  peak radiation temperature on the capsule is (284.7, 286.7) eV in the (SRS at
  lens, inline SRS) simulations, and occurs at 14.6 ns in both}\BibitemShut
  {NoStop}%
\bibitem [{\citenamefont {Michel}\ \emph {et~al.}(2013)\citenamefont {Michel},
  \citenamefont {Rozmus}, \citenamefont {Williams}, \citenamefont {Divol},
  \citenamefont {Berger}, \citenamefont {Glenzer},\ and\ \citenamefont
  {Callahan}}]{michel-heating-pop-2013}%
  \BibitemOpen
  \bibfield  {author} {\bibinfo {author} {\bibfnamefont {P.}~\bibnamefont
  {Michel}}, \bibinfo {author} {\bibfnamefont {W.}~\bibnamefont {Rozmus}},
  \bibinfo {author} {\bibfnamefont {E.~A.}\ \bibnamefont {Williams}}, \bibinfo
  {author} {\bibfnamefont {L.}~\bibnamefont {Divol}}, \bibinfo {author}
  {\bibfnamefont {R.~L.}\ \bibnamefont {Berger}}, \bibinfo {author}
  {\bibfnamefont {S.~H.}\ \bibnamefont {Glenzer}}, \ and\ \bibinfo {author}
  {\bibfnamefont {D.~A.}\ \bibnamefont {Callahan}},\ }\href
  {http://scitation.aip.org/content/aip/journal/pop/20/5/10.1063/1.4802828}
  {\bibfield  {journal} {\bibinfo  {journal} {Phys. Plasmas}\ }\textbf
  {\bibinfo {volume} {20}},\ \bibinfo {eid} {056308} (\bibinfo {year}
  {2013})}\BibitemShut {NoStop}%
\end{thebibliography}

%

\end{document}